\documentclass[10pt]{article}
\usepackage[utf8]{inputenc}
\pdfoutput=1
\usepackage{amssymb,amsmath,mathrsfs,enumerate}
\usepackage{graphicx,rotate,multicol}
\usepackage{float}
\usepackage{subfig}
\usepackage[margin=10pt,labelfont=bf]{caption}
\usepackage{cite}
\usepackage[colorlinks=true,
            linkcolor=blue,
            urlcolor=blue,
            citecolor=teal]{hyperref}

\usepackage{color}
\definecolor{ourcolor}{rgb}{0.7, 0.25, 0.05}
\usepackage{tikz,braket}
\long\def\rpl#1!!#2!!{\textcolor{red}{#1} \textcolor{blue}{#2}}

\let\tilde=\widetilde

\let\bar=\overline

\def \order(#1){{\mathcal O} \left(#1 \right)}

\evensidemargin=\oddsidemargin
\allowdisplaybreaks

\title	{\color{ourcolor} \LARGE \bf 750 GeV Resonance in the Dark Left-Right Model} 

\author {\sf Ujjal Kumar Dey,$^{a,}$\footnote{ujjaldey@prl.res.in} \hspace{4pt}  Subhendra Mohanty,$^{a,}$\footnote{mohanty@prl.res.in} \hspace{4pt} Gaurav Tomar$^{a,b}$\footnote{tomar@prl.res.in} \\[10pt]
\small\em $^a$ Theoretical Physics Division, 
		Physical Research Laboratory,
		Navrangpura, Ahmedabad 380009, India\\
\small\em $^b$Indian Institute of Technology, Gandhinagar 382424, India\\
}

\date{}

\begin{document}

\maketitle	

\begin{abstract}
We explain the $750$ GeV diphoton resonance in the context of the dark left-right symmetric model. A global symmetry in this model,  stabilizes the dark matter and ensures that the scalar couples dominantly to gluons and photons. The branching fraction of the scalar to diphoton is large as a consequence of the symmetries of the theory. The benchmark values assumed to fit the diphoton signal also give the correct relic density of dark matter and muon $(g-2)$. A specific prediction of this model is that the dark matter has a mass of 200 GeV.
\end{abstract}

~~~~Keywords: Left-right symmetry; Dark matter; Muon magnetic moment; Scalar resonance

\bigskip

\section{Introduction}
\label{sec:intro}
A new diphoton excess has been reported by the ATLAS \cite{atl-dp} and CMS collaborations \cite{cms-dp} in their analysis of $\sqrt{s}=13$ TeV $pp$ collision.
The cross-section of $\sigma(pp \rightarrow \gamma \gamma)$ for the $13$ TeV run has been given as the excess of the cross-section in $\gamma\gamma$ channel can be estimated as $(6 \pm 3)$ fb (CMS \cite{cms-dp}) and $(10 \pm 3)$ fb (ATLAS \cite{atl-dp}). The excess has taken a wide attention in the community and a wide range of models has been proposed to explain this excess.
Since the two photon decay can occur from a spin-0 or spin-2 particle resonance, the models considered of elementary scalars~\cite{Agrawal2015, Aloni2015, Angelescu2016, Backovic2015, Becirevic2015, Buttazzo2016, Cao2015, Chao2015, Csaki2016, Curtin2016, Demidov2015, DiChiara2015, Dutta2015, Ellis2015, Falkowski2015, Franceschini2015, Gupta2015, Knapen2015, Kobakhidze2015, Low2015, Mambrini2015, Martinez2015, McDermott2015, Petersson2015, Chakrabortty2015, Fichet2015, Ahmed2015, Cox2015, Chao2015a, Bi2015, Bardhan2015, Heckman2015, Barducci2015, Cao2015a, Ding2015, Han2015b, Han2015a, Alves2015, Antipin2015}, composite pseudo-Nambu-Goldstone bosons \cite{Bai2015, Bellazzini2015, Bian2015, Harigaya2015, Matsuzaki2015, Nakai2015, No2015, Higaki2016, Molinaro2015, Pilaftsis2016}, Kaluza-Klein graviton \cite{Arun2015, Han2015}  or cascade decays of heavier particles \cite{Cho2015, Huang2015}. The effect of 750 GeV resonance on vacuum stability is studied in \cite{Dhuria2015, Hamada2015, Salvio2016}. Previously, the resonant production of (pseudo-)scalars coupled to two photons and gluons in the mass region from ~30 GeV to 2 TeV at the LHC has been studied in~\cite{Jaeckel2013}\footnote{ A comprehensive check of 750 GeV diphoton resonance models has been performed by implementing them in the SARAH \cite{Staub2008, Staub2014} package by Staub et al. \cite{Staub2016}}.  
The resonance must be produced from $gg$ or $qq$ vertices and decay to $\gamma\gamma$ via exotic fermions, gauge bosons or scalars in the loop. It would be of interest to examine the possibility of a $750$ GeV resonance in an existing well motivated particle physics model. A well studied generalization of the Standard Model (SM) is the left-right model with the gauge group $SU(2)_L\times SU(2)_R\times U(1)$ and the same fermions content as the SM. In the left-right model, the right handed neutrino would have been a good candidate as a Dark Matter (DM) if it could be made stable against decays via $W^\pm_R$. To ensure the stability of the dark matter, an extra global symmetry $S$ is imposed in the so called dark left-right gauge model (DLRM) \cite{Khalil:2009nb,Aranda2010}. The spontaneous breaking of $SU(2)_R\times S$ is achieved in such a way as to leave the combination $\tilde{L}\equiv S-T_{3R}$ unbroken. The fermion sector is extended by adding two quarks. The generalized lepton number is $\tilde L=1$ for the SM leptons and $\tilde L=0$ for all other SM particles. To achieve this pattern of symmetry breaking and give masses to the fermions, one adds two doublets to the usual higgs sector of the left-right model. In this model the right-handed gauge boson $W^\pm_R$ has $\tilde L=1$ and this prevent the decay of the right-handed neutrino $(n_R)$ via $W_R^{\pm}$ and makes the lightest right-handed neutrino a candidate for dark matter \cite{Khalil:2009nb,Aranda2010,Basak2015}. It has also been shown that the charged scalars of the triplet higgs of the DLRM model can give an adequate contributions to the muon $(g-2)$.
In this paper, we explain the $750$ GeV diphoton excess \cite{atl-dp, cms-dp} in the DLRM model. The neutral component ($\phi_{R}^{0}$) of the right-handed higgs doublet $\Phi_R$ is interpreted as the resonance. It can couple to gluons via the exotic quarks $(x)$ and to photons via the same quark, $W^\pm_R$ and right-handed charged scalars $(\Delta^\pm_R,\Delta^{\pm\pm}_R)$. Interestingly, in this model $\phi^0_R~(750~\mbox{GeV})$ dominantly decays into gluons and photons pairs and this explain the large value of $\Gamma(\phi^0_R\rightarrow gg)\Gamma(\phi^0_R\rightarrow \gamma\gamma)/\Gamma_{\mbox{tot}}$ needed to explain the diphoton signal \cite{atl-dp, cms-dp}. 
The cross-section $\sigma (pp \rightarrow \gamma\gamma)$ depends upon the $x$ Yukawa coupling with $\phi^0_R$, the charged higgs masses $m_{\Delta_{R}^\pm},m_{\Delta_{R}^{\pm\pm}}$, and the quartic couplings $f_1,f_2$ of $\Phi_R \Phi_R \Delta_R \Delta_R$ interaction. There is a negative contribution to the $\phi^0_R \rightarrow \gamma\gamma$ amplitude from the $W^\pm_R$ loop, but this is small due to the large $W^\pm_R$ mass. We scan the parameter space of these couplings and masses, which give the cross-section
$\sigma(pp \rightarrow \phi^0_R \rightarrow \gamma\gamma)=$ 3-13 fb. We compute the muon ($g-2$) where the $\Delta_{R}^{\pm}$ and $\Delta_{R}^{\pm\pm}$ contributes in the loop. We find that by taking the masses of the charged scalars $m_{\Delta_{R}^{\pm}} \simeq m_{\Delta_{R}^{\pm\pm}} = 530$ GeV and their Yukawas with $\mu$ of the order of unity we get the required $\Delta a_{\mu}$ of $\sim 3\times 10^{-9}$. We identify the second generation $n_R$ as the dark matter. With these parameters fixed we find that the correct relic density is obtained by taking the dark matter mass $200$ GeV. 
The paper is organized as follows: in Sec.~\ref{sec:dlrm} we described the dark left-right model. In Sec.~\ref{sec:h1decay} we explain the observed diphoton signal in this model. The dark matter aspects and the constraints from muon ($g-2$) are discussed in Sec.~\ref{sec:dmmug} and finally we conclude in Sec.~\ref{sec:concl}.
\section{Dark Left-Right Model}
\label{sec:dlrm}
We consider the dark left-right gauge model (DLRM) \cite{Khalil:2009nb,Aranda2010} with gauge a group $SU(3)_C \times SU(2)_R \times SU(2)_L \times U(1)$ of the usual left-right model. In DLRM, there is an extra $U(1)$ global symmetry $S$ to ensure that the spontaneous breaking of  $SU(2)_R \times S$ leaves the global symmetry $\tilde{L}=S-T_{3R}$ unbroken. 
Here $\tilde{L}$ is identified as generalized lepton number, with $\tilde{L}=1$ for the SM leptons, and $\tilde{L}=0$ for all other SM particles. In this model $W^\pm_R$ also carries a lepton number, which ensures the stability of dark matter. 
In addition to the SM fermions (shown in Table~\ref{tab:ferm}), model contains new neutral lepton $(n_R)$ and exotic quark $(x_R)$. The model also have a new $SU(2)_{L,R}$ singlet quark $(x_L)$. The exotic quark $(x_L,x_R)$, $n_R$ carry the generalized lepton number $\tilde L=1$. This ensures that the Yukawa terms can be written without breaking the global symmetry $S$. 
\begin{table}[!htbp]
\centering
\begin{tabular}{|c|c|c|c|}
\hline
Fermion                   & $SU(3)_{c}\otimes SU(2)_{L}\otimes SU(2)_{R}\otimes U(1)$ & $S$   & $L$     \\ \hline \hline
$\Psi_{L} = (\nu, e)_{L}$ & $(1,2,1,-1/2)$                                            & $1$   & $(1,1)$ \\ \hline
$\Psi_{R} = (n, e)_{R}$   & $(1,1,2,-1/2)$                                            & $1/2$ & $(0,1)$ \\ \hline
$Q_{L} = (u, d)_{L}$      & $(3,2,1,1/6)$                                             & $0$   & $(0,0)$ \\ \hline
$Q_{R} = (u, x)_{R}$      & $(3,1,2,1/6)$                                             & $1/2$ & $(0,1)$ \\ \hline
$d_{R}$                   & $(3,1,1,-1/3)$                                            & $0$   & $0$     \\ \hline
$x_{L}$                   & $(3,1,1,-1/3)$                                            & $1$   & $1$     \\ \hline
\end{tabular}
\caption{Fermion content of DLRM model.}
\label{tab:ferm}
\end{table}
The scalar sector of DLRM consists of a bi-doublet $(\Phi)$, two triplets $(\Delta_L, \Delta_R)$ and two doublets ($\Phi_L, \Phi_R$) such as,
\begin{center}
            $ \Phi=\begin{pmatrix}
                \phi_1^0 & \phi_2^+\\
                \phi_1^- &  \phi_2^0
               \end{pmatrix} $,
             $\Phi_{L}=\begin{pmatrix}
                   \phi_{L}^+\\
                    \phi_{L}^0
                  \end{pmatrix} $,
                  $\Phi_{R}=\begin{pmatrix}
                   \phi_{R}^+\\
                    \phi_{R}^0
                  \end{pmatrix} $ and
            $ \Delta_{L,R}=\begin{pmatrix}
                \frac{\Delta_{L,R}^{+}}{\sqrt{2}} & \Delta_{L,R}^{++}\\
                \Delta_{L,R}^{0} &  -\frac{\Delta_{L,R}^{+}}{\sqrt{2}}
               \end{pmatrix} $,
\end{center}
The quantum numbers of scalars under the DLRM gauge group and $S$ are listed in Table~\ref{tab:scalar}. 
\begin{table}[!htbp]
\centering
\begin{tabular}{|c|c|c|}
\hline
Scalar         & $SU(3)_{c}\otimes SU(2)_{L}\otimes SU(2)_{R}\otimes U(1)$ & $S$    \\ \hline \hline
$\Phi$         & $(1,2,2,0)$                                               & $1/2$  \\ \hline
$\tilde{\Phi}$ & $(1,2,2,0)$                                               & $-1/2$ \\ \hline
$\Phi_{L}$     & $(1,2,1,1/2)$                                             & 0      \\ \hline
$\Phi_{R}$     & $(1,1,2,1/2)$                                             & $-1/2$ \\ \hline
$\Delta_{L}$   & $(1,3,1,1)$                                               & $-2$   \\ \hline
$\Delta_{R}$   & $(1,1,3,1)$                                               & $-1$   \\ \hline
\end{tabular}
\caption{Scalar content of DLRM model. Note that $\tilde{\Phi} = \sigma_{2}\Phi^{\ast}\sigma_{2}$.}
\label{tab:scalar}
\end{table}
As a result of global symmetry $S$, not all of the Yukawa terms are allowed. The allowed Yukawa terms are:
\begin{align}
 \bar \psi_L \Phi \psi_R,~\bar Q_L \tilde \Phi Q_R,~\bar Q_L \Phi_L d_R,~\bar Q_R \Phi_R x_L,~\psi_L \psi_L \Delta_L,~\mbox{and}~\psi_R \psi_R \Delta_R,
 \label{yukawa}
\end{align}
whereas $S$ forbids the Yukawa terms like: $\bar \psi_L \tilde \Phi \psi_R,~\bar Q_L \Phi Q_R$ and $\bar x_L d_R$. So the masses of charged leptons and up-quark come from the vacuum expectation value (vev) $v_2=\langle \phi^0_2 \rangle$, mass of the down-quark comes from  $v_3=\langle \phi^0_L \rangle$, mass of the exotic quark comes from $v_4=\langle \phi^0_R \rangle$, mass of neutrino comes from $v_5=\langle \Delta^0_L \rangle$, and mass of $n_R$ comes from $v_6=\langle \Delta^0_R \rangle$. The flavor-changing neutral currents will be absent at the tree level as a result of considered model structure.\\
The scalar potential with terms allowed by $S$ is of the form,
\begin{align}
 V =& \left(m_1^2 \Phi^\dag \Phi + m_3^2 \Phi_L^\dag \Phi_L+m_4^2 \Phi_R^\dag \Phi_R+m_5^2 \Delta_L^\dag \Delta_L
 +m_6^2 \textrm{Tr}(\Delta_R^\dag \Delta_R)\right) +\nonumber \\ 
  & ~~\Phi_R^\dag \Delta_R \tilde{\Phi}_{R}+ \Phi_L^\dag \Phi \Phi_R+\textrm{Tr}(\tilde{\Phi}^\dag \Delta_L \Phi \Delta_R^\dag)+\text{quartic-terms}.
\end{align}
The higgs potential leaves the combination $\tilde L=S-T_{3R}$ unbroken giving rise to solution $v_1=\langle \phi^0_1 \rangle=0$. We will focus on the $SU(2)_R$ higgs structure of DLRM and consider the most general higgs potential containing $\Phi_R$ and $\Delta_R$ \cite{Aranda2010},
\begin{eqnarray}
 V_R &=& m^2_4 \Phi^\dagger_R \Phi_R + m^2_6 \textrm{Tr}(\Delta^\dagger_R \Delta_R) + \frac{1}{2} \lambda_1 (\Phi^\dagger_R \Phi_R)^2 
      + \frac{1}{2} \lambda_2 [\textrm{Tr}(\Delta^\dagger_R \Delta_R)]^2 \nonumber \\
      && + \frac{1}{4} \lambda_3 \textrm{Tr}(\Delta^\dagger_R \Delta_R-\Delta_R \Delta^\dagger_R)^2 + f_1 (\Phi^\dagger_R \Phi_R) \textrm{Tr}(\Delta^\dagger_R \Delta_R)\nonumber \\
      && + f_2 \Phi^\dagger_R (\Delta^\dagger_R \Delta_R-\Delta_R \Delta^\dagger_R)\Phi_R + \mu (\Phi^\dagger_R \Delta_R \tilde \Phi_R + \tilde \Phi^\dagger_R \Delta^\dagger_R \Phi_R).
\end{eqnarray}
After the electroweak symmetry breaking, the mass matrix of the CP-even neutral scalars contains off-diagonal terms and give as,
\begin{center}
$\mathcal{M}^2\left( \textrm{Re}\left(\phi^0_R\right), \textrm{Re}\left(\Delta^0_R\right)\right) =\begin{pmatrix}
                2\lambda_1 v^2_4 & 2(f_1-f_2)v_4 v_6 + 2\mu v_4\\
                2(f_1-f_2)v_4 v_6 + 2\mu v_4 &  2 (\lambda_2+\lambda_3)v^2_6-\mu v^2_4/v_6
               \end{pmatrix}$.
\end{center}
If we assume a small mixing between $\phi^0_R$ and $\Delta^0_R$ by taking $f_1=f_2$ and $\mu\ll \lambda_1 v_4$, then the lowest mass eigenstate is,
\begin{eqnarray}
  m^2_{H_1} \approx 2\lambda_1 v^2_4,
\end{eqnarray}
which we identify as $750$ GeV resonance observed at LHC.
We will now discuss the relevant interactions of $H_{1} = \phi_{R}^{0}$, which are responsible for various decay modes. 
\begin{figure}[!htbp]
\begin{center}
  \subfloat[\label{sf:ggh1}]{
   \includegraphics[scale=0.4]{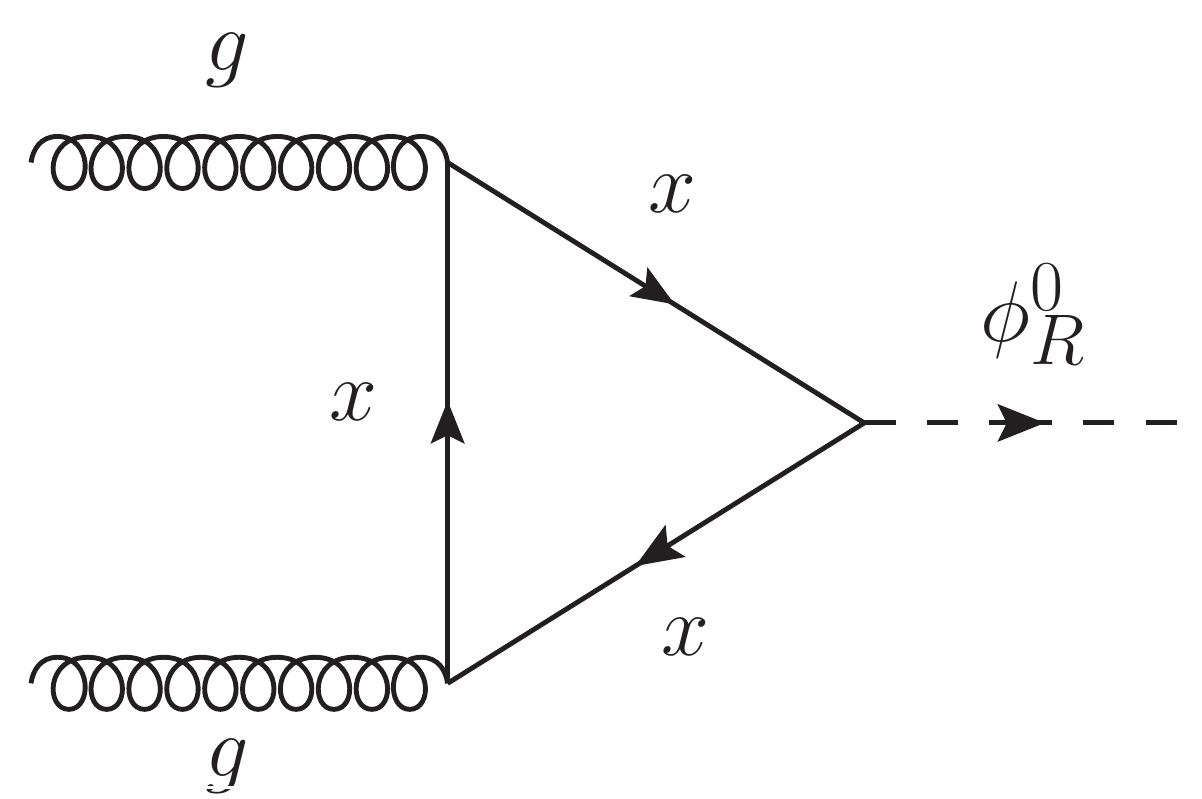}}\\
  \subfloat[\label{sf:yyh1}]{
   \includegraphics[scale=0.3]{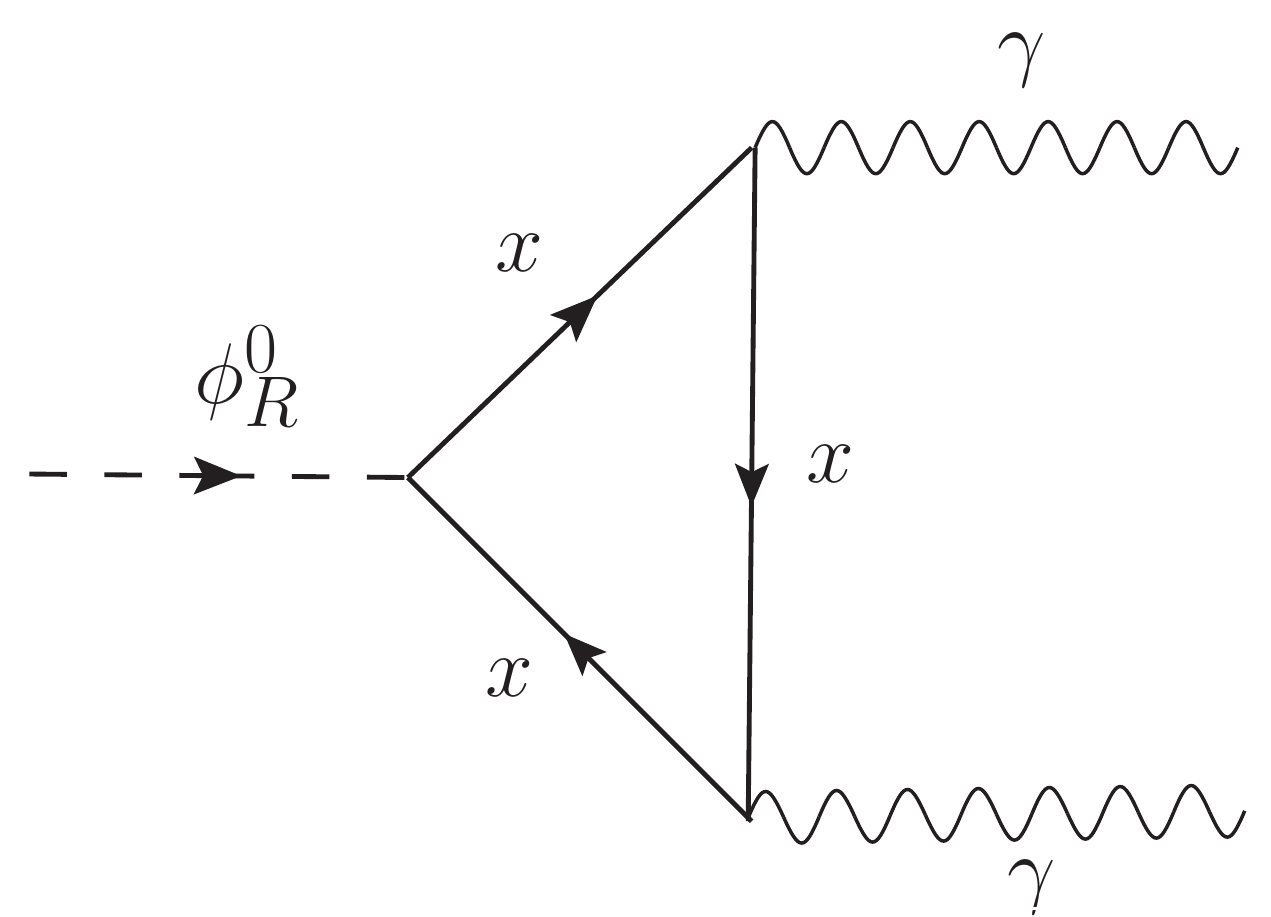}~~
   \includegraphics[scale=0.3]{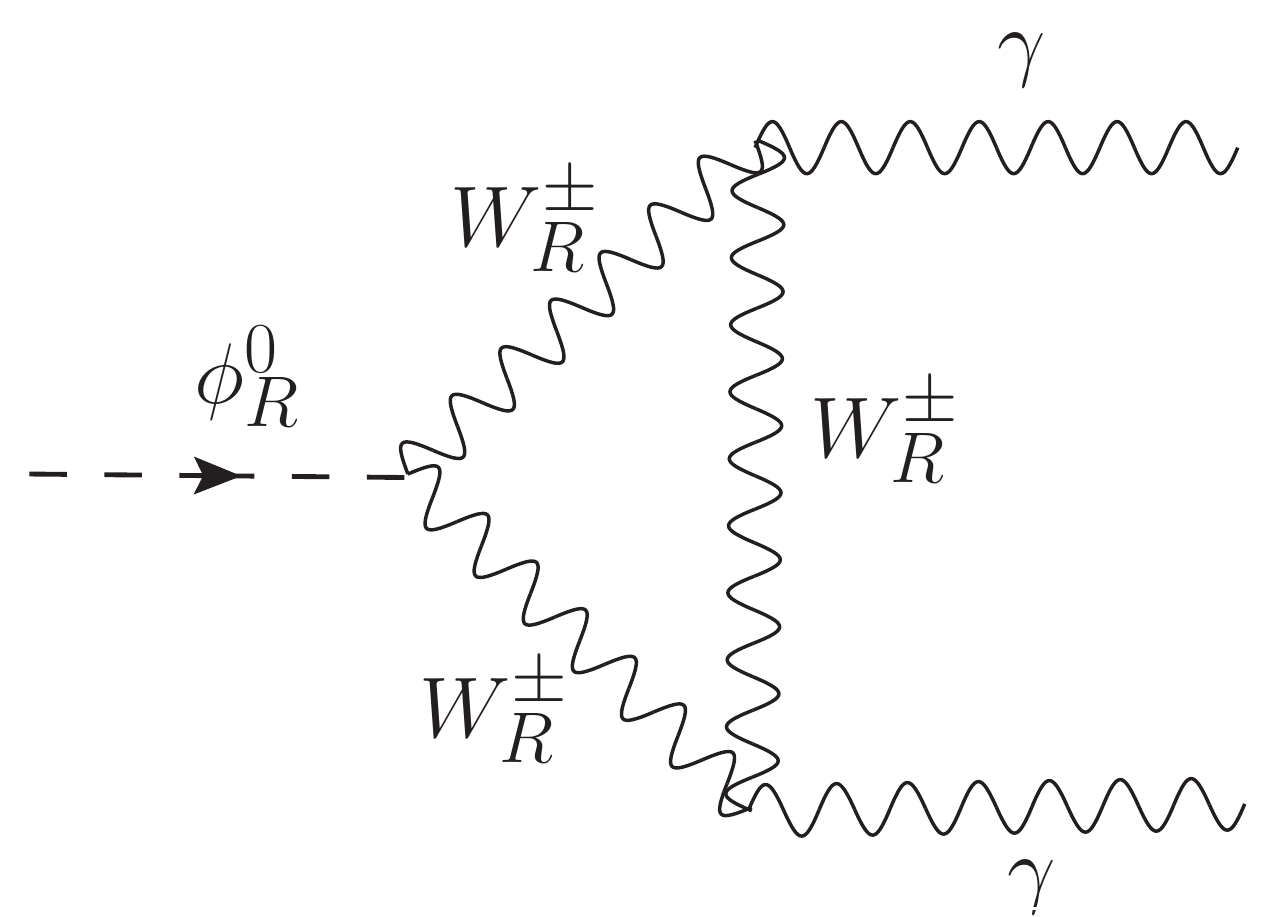}~~
   \includegraphics[scale=0.3]{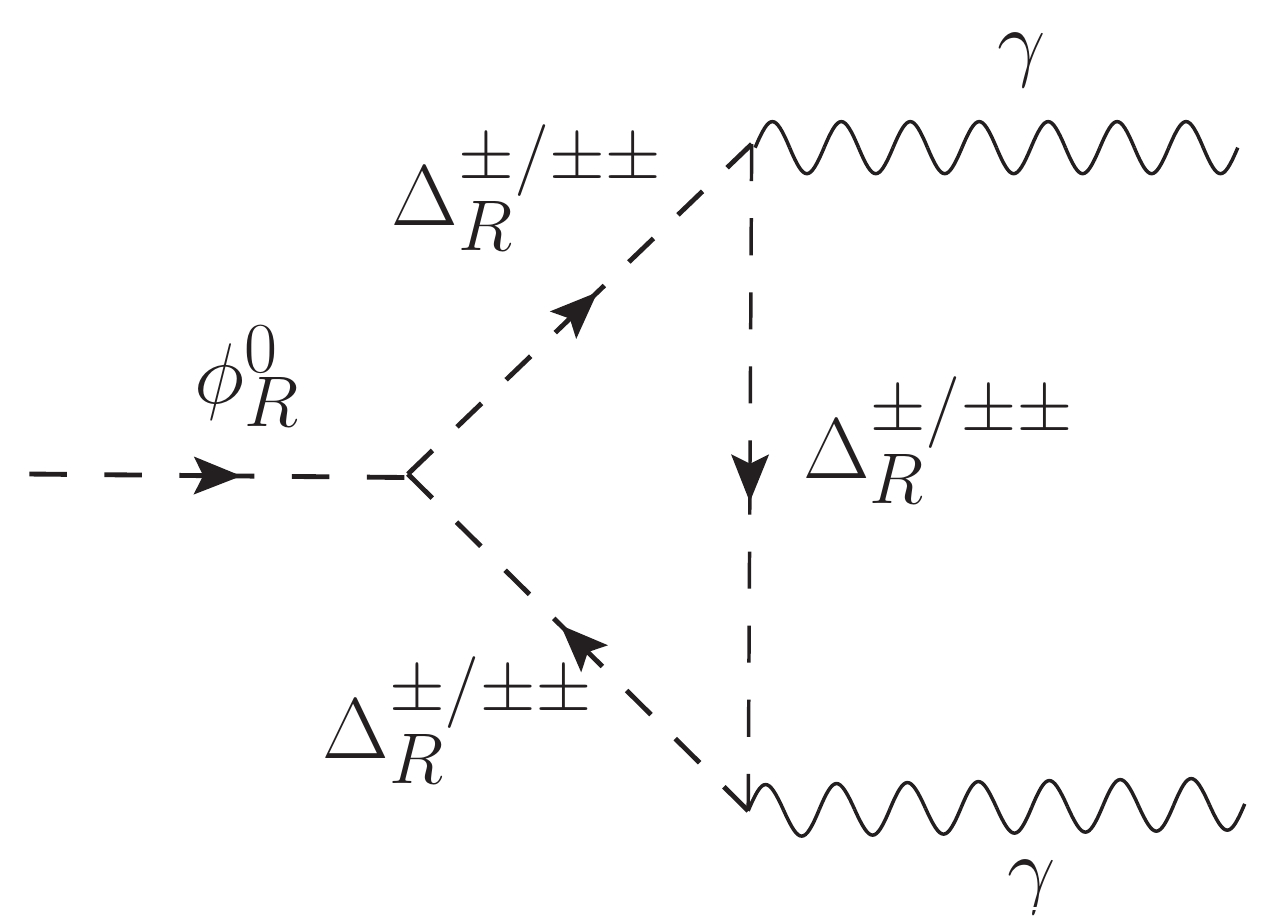}}
   \caption{Feynman diagrams for the production and decay of $\phi_{R}^{0}$.}
\label{f:ggh1}
\end{center}
\end{figure}

There exists exotic quark $(x)$ in the model, the loop induced process of which can produce $\phi^0_R$ through gluon fusion. The $\phi^0_R$ can subsequently decay to photons (as shown in Fig.~\ref{f:ggh1}) giving rise to the observed diphoton signal. The relevant Yukawa interaction is $\bar Q_R \Phi_R x_L$, which translates to $ \bar u_R \phi^+_R x_L + \bar x_R \phi^0_R x_L $.
The resonance $\phi^0_R$ can also decay through $W^{\pm}_R,~\Delta^\pm_R$ and $\Delta^{\pm\pm}_R$, by the following interactions,
\begin{equation}
  g_R~\phi^0_R {W^\mu_R} W_{R\nu},~2 v_4 f_1 \Delta_{R}^+ \Delta_{R}^- \phi^0_R,~\mbox{and}~~ 2 v_4 (f_1+f_2) \Delta_{R}^{++} \Delta_{R}^{--} \phi^0_R.
\end{equation}
\section{Diphoton Signal}
\label{sec:h1decay}
The diphoton signal cross-section, at the LHC for the proton center-of-mass energy $\sqrt{s}$, is given by,
\begin{align}
\sigma(pp\to \phi^0_R \to \gamma \gamma) = \frac{1}{m_{H_1}s\Gamma_{\rm tot}}
         [c_{gg} \Gamma(\phi^0_R \to gg)]
         \Gamma(\phi^0_R \to \gamma \gamma),
         \label{meq}
\end{align}
where $c_{gg}$ represents the parton integral,
\begin{align}
c_{gg} = \frac{\pi^{2}}{8}\int_{m_{H_{1}}^{2}/s}^{1}
         \frac{dz}{z}g(z)g\left(\frac{m_{H_{1}}^{2}}{zs}\right).
\end{align}
  For $\sqrt{s} = 13$ TeV and $m_{H_{1}} = 750$ GeV, one can get $c_{gg}$ to be $\sim$ 2137 \cite{Franceschini2015}. Now, within the 1$\sigma$ error of CMS and ATLAS data the cross section $\sigma\in [3,13]$ fb. In our analysis we use this range and calculate the allowed parameter space, consistent with this cross section measurement, allowed in our model. In this analysis we fix the mass of $\phi_{R}^{0}$, i.e., $m_{H_{1}}$ at 750 GeV.
We compute the decay width $\Gamma (\phi^0_R \to gg)$ with gluons, which has contribution from the new quark $x$, as shown in Fig.~\ref{sf:ggh1}. We also compute the  $\Gamma (\phi^0_R \to \gamma\gamma)$ decay width; the relevant diagrams are shown in Fig.~\ref{sf:yyh1}. The total $\phi^0_R$ decay width is $\Gamma_{\rm tot} =\Gamma (\phi^0_R \to gg)+\Gamma (\phi^0_R \to \gamma\gamma)$, since $\phi^0_R$ does not have couplings with any SM fermions, $W^\pm$ and $Z$. It couples to scalars $\Delta^{\pm/\pm \pm}_{R}$, but we assume the masses of these scalars in the range such that the decay of $\phi_{R}^{0}$ to these scalars are kinematically forbidden. The contribution will come from $x$, $W_{R}^{\pm}$, $\Delta_{R}^{\pm}$ and $\Delta_{R}^{\pm \pm}$, see Fig.~\ref{f:ggh1}.
 \begin{figure}[!t]
  \begin{center}	
   \subfloat[\label{v4mS}]{
   \includegraphics[scale=0.45]{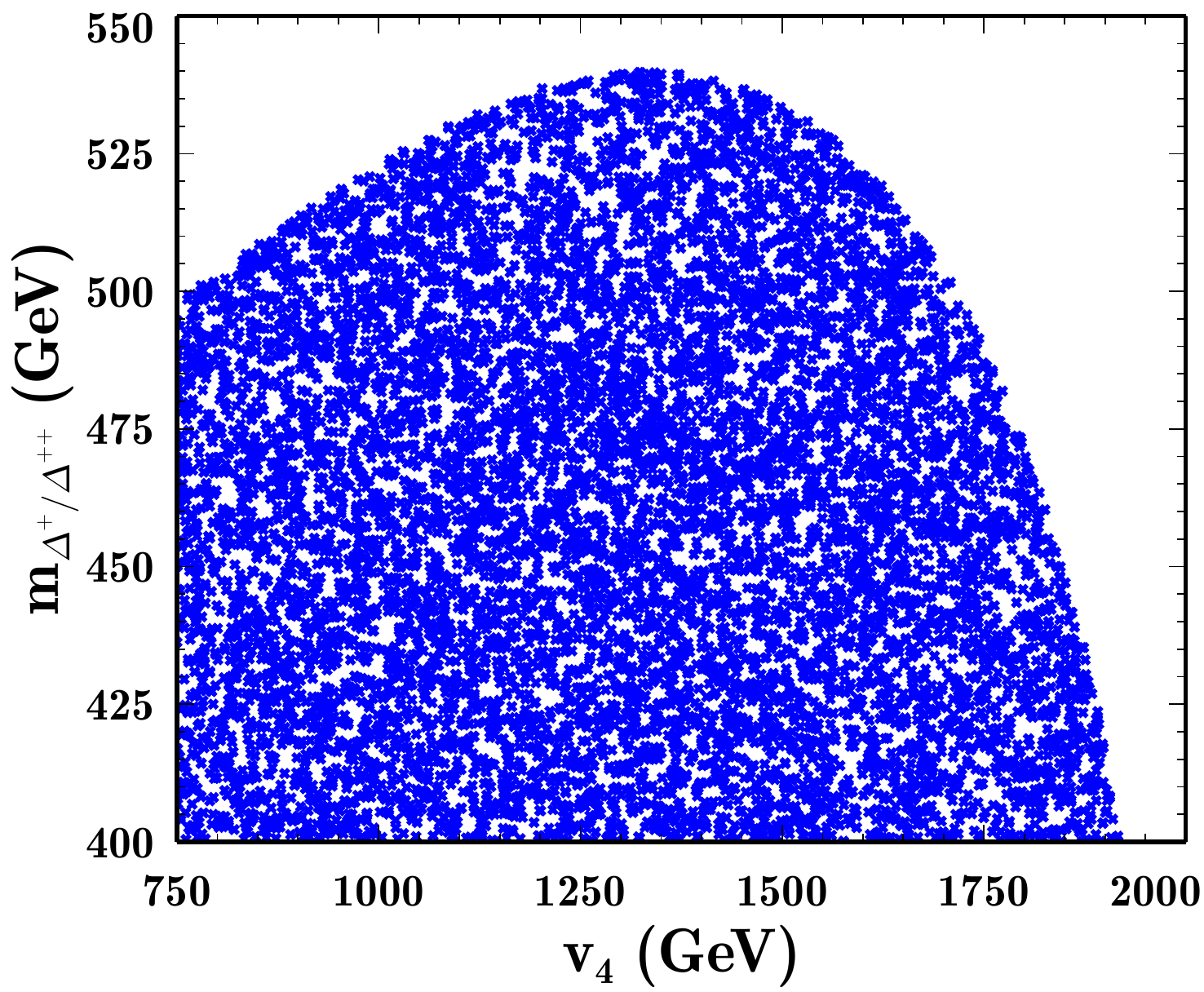}}~~~~~~
   \subfloat[\label{v4gS}]{
   \includegraphics[scale=0.45]{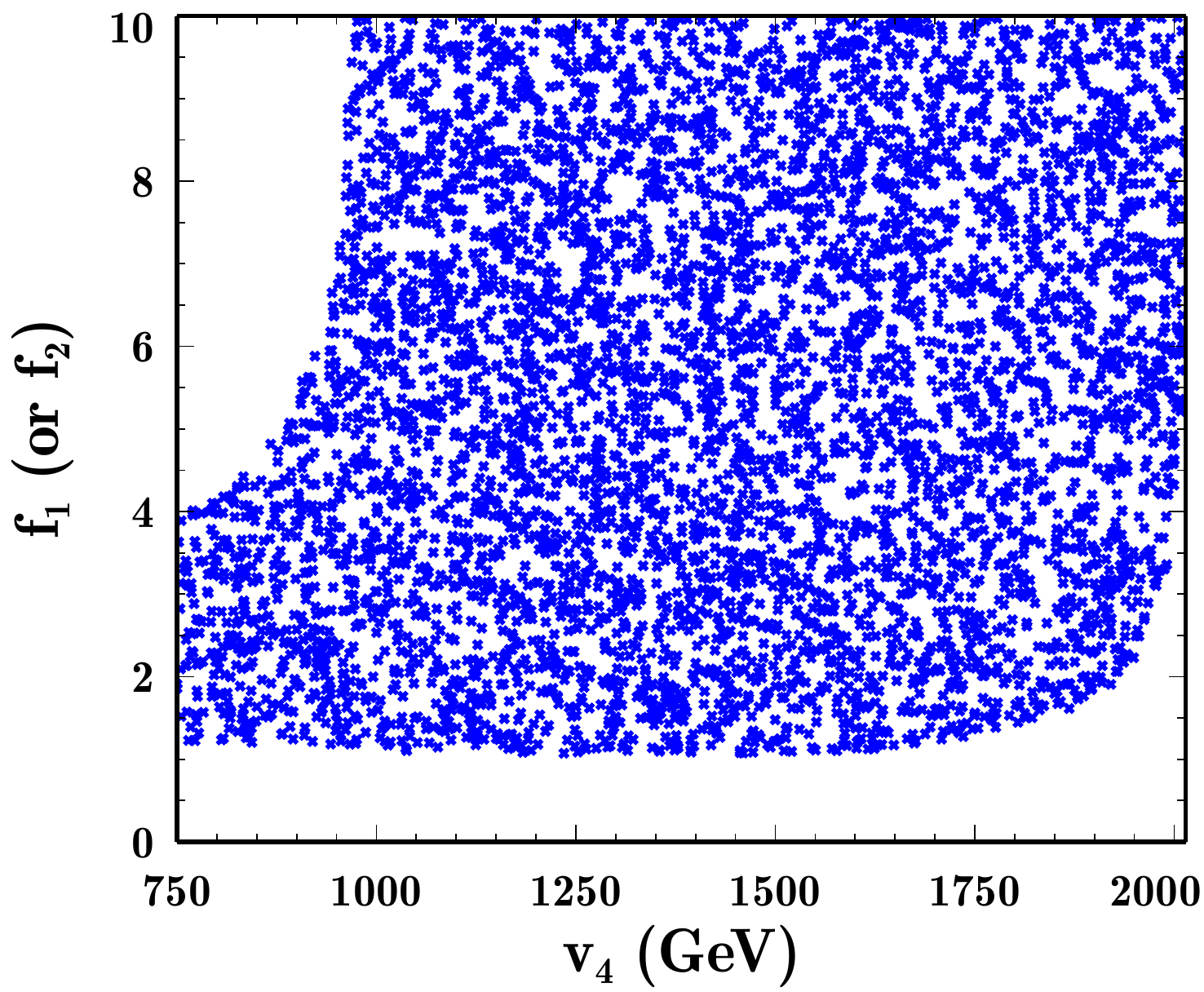}}
 \caption{\it Allowed parameter space which results in the correct $pp\to \gamma \gamma$ cross section. We take $m_{W_{R}} = 1.8$ TeV and $g_{R} = 0.1$. In the left panel we show the allowed parameter space of the vev $v_{4}$ and the charged scalar masses. For simplicity we assume $f_{1} = f_{2}=1$. The right panel shows the parameter region of $v_{4}$ and the quartic couplings $f_{1}$ and $f_{2}$. We assumed $m_{\Delta^{\pm}_{R}} = m_{\Delta^{\pm \pm}_{R}}=530$ GeV.}
 \label{f:plots}
 \end{center} 
\end{figure}
In this context we would like to mention the experimental constraints on the masses of $W_{R}^{\pm},~\Delta^\pm_R$ and $\Delta^{\pm\pm}_R$. In DLRM $M_{Z_R}$ and $M_{W^\pm_R}$ are related as \cite{Khalil:2009nb},
\begin{equation}
 M^2_{W_R} > \frac{(1-2 s_w^2)}{2(1-s_w^2)}~M^2_{Z_R} + \frac{s_w^2}{2(1-s_w^2)^2}~M^2_{W_L},
 \label{wrm}
\end{equation}
where zero $Z-Z_R$ mixing is assumed. $Z_R$ decays into SM fermions; $Z_R\rightarrow \ell^+\ell^-$ have been reported by CMS (ATLAS) collaboration, which put $M_{Z_R}> 2.6~\mbox{TeV}~(2.9~\mbox{TeV})$ \cite{Chatrchyan2013}. After using Eq.~\ref{wrm}, these bounds translate for $W_{R}^{\pm}$
giving $M_{W_R}\textgreater 1.5~~\mbox{TeV}~(1.7~\mbox{TeV})$. The $W^\pm_R$ diagram contributes with a negative sign, but by taking $M_{W_R}=1.8$ TeV we find that  $W^\pm_R$ diagram has a smaller amplitude in compare to other diagrams of Fig.~\ref{f:ggh1}.
The doubly charged scalar ($\Delta^{\pm\pm}_R$) dominantly decays into same sign dileptons and is constrained by CMS (ATLAS) collaboration, which exclude $m_{\Delta^{\pm\pm}}$ below 445 GeV (409 GeV) and 457 GeV (398 GeV) for $e^\pm e^\pm$ and $\mu^\pm \mu^\pm$ channels respectively \cite{cmsl,Aad2012}. The mass of singly charged scalar below 600 GeV (assuming 100\% branching ratio in the $\tau\nu_{\tau}$ channel) is ruled out at $95\%$ confidence level \cite{Chatrchyan2012,atlass}. But in DLRM model $\Delta^\pm_R$ not only couples with $\tau^\pm$, but also with $e^\pm$ and $\mu^\pm$, so this bound is relaxed.
In Fig.~\ref{f:plots} we show the allowed parameter space of our model which gives $\sigma (pp \to \gamma\gamma) \sim 3-13$ fb, consistent with the observed diphoton signal cross section. In these plots we have taken $m_{\Delta^{\pm}_{R}} = m_{\Delta^{\pm \pm}_{R}}$ and $f_{1} = f_{2}$. From the figures it is evident that $v_{4}\gtrsim 1900$ GeV is not suitable for explaining the diphoton excess. For example,  taking $f_{1} = f_{2} = 1$, $v_{4} = 1.2$ TeV and $m_{\Delta^{\pm}_{R}} = m_{\Delta^{\pm \pm}_{R}}=460$ GeV we found the $\sigma(pp\to \gamma \gamma)\sim 4.2$ fb which lies in the range of observed diphoton cross section; also for these set of parameters the decay width to $gg$ is $1.9\times 10^{-3}$ GeV  and to $\gamma \gamma$ is $1.2\times 10^{-3}$ GeV. Note that the 750 GeV scalar $\phi_{R}^{0}$, in this model, can also decay, via loop, to $Z\gamma$ and $ZZ$, but these decays are smaller than the diphoton decay by the factors $\sim 1/3$ and $1/9$ respectively and we have included these in our analysis. At this point it is worth-mentioning that this model can not reproduce the broad decay width mildly favoured by ATLAS data. In our subsequent calculation for dark matter relic density and muon $(g-2)$, we take  $m_{\Delta^{\pm}_{R}} = m_{\Delta^{\pm \pm}_{R}}=530$ GeV, which is consistent with the diphoton signal.

\section{Dark Matter and muon $(g-2)$}
\label{sec:dmmug}

\subsection{Dark Matter Relic Density}
\label{sbsc:dmrelic}
As an consequence of global symmetry $S$, the Yukawa term $\psi_L \tilde \Phi \psi_R$ is forbidden in DLRM, which basically connects $\nu_L$ with $n_R$. This implies that $\nu_L$ and $n_R$ are not Dirac mass partner and the lightest $n_R$ (out of three possible generation) can be identified as a viable dark matter candidate. 
We identify the second generation $n_R~(n^\mu_R)$ as dark matter (we will call it $\chi$), because then it can be related to the discrepancy of muon $(g-2)$. This idea is explored in detail in \cite{Basak2015}.
The mass of dark matter generates through the Yukawa term $\Psi_R\Psi_R \Delta_R$. We assume that $m_{\Delta_{R}^\pm}$ is lighter in comparison to $W_R^{\pm}, Z_R$ gauge boson masses. Therefore, the dominant annihilation of dark matter is given by a $t$-channel $\chi\chi\rightarrow \mu^+\mu^-$ through $\Delta_{R}^\pm$.\\
 \begin{figure}[!htbp]
  \begin{center}	
   \includegraphics[height=7cm,width=8cm]{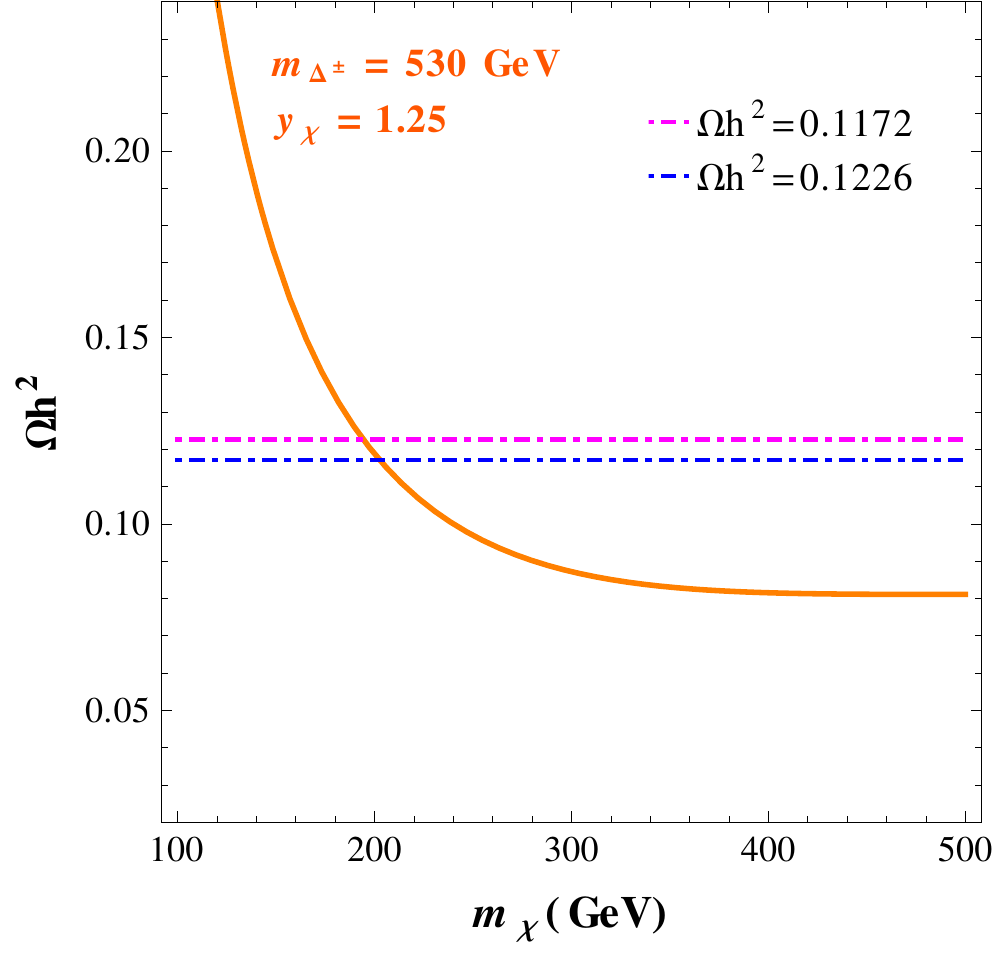}~~~~~~
    \caption{\it The relic abundance as a function of dark matter mass for $y_\chi=1.25$ and $m_{\Delta_{R}^{\pm}} = 530$ GeV. The relic density experimental value $\Omega h^2=0.1199\pm 0.0027$~\cite{Ade2014} is shown by straight lines.}
 \label{relic:plots}
 \end{center} 
\end{figure}
After using the partial-wave expansion, the thermally averaged annihilation cross-section can be written as $\langle \sigma v \rangle \simeq a + 6b/x_f$, where $a$ and $b$ are the $s$ and $p$-wave contributions respectively. The $s$-wave contribution is helicity suppressed and given as \cite{Ellis1998,Nihei2002},
\begin{equation}
 a \simeq \frac{y_{\chi}^4}{32\pi m_\chi^2}\frac{m_f^2}{m_\chi^2}\frac{1}{(1+w)^2}
\end{equation}
and the $p$-wave contribution is \cite{Cao2009}, 
\begin{equation}
 b \simeq \frac{y_{\chi}^4}{48\pi m_\chi^2}\frac{(1+w^2)}{(1+w)^4}
\end{equation}
where $m_{f}$ is the mass of the final state fermions and $y_\chi$ is the Yukawa type coupling between $\chi,~\mu^-$ and $\Delta_{R}^+$. Here the ratio of charged scalar mass with dark matter $\chi$ is defined as, $w=m_{\Delta_R^+}^2/m_\chi^2$. It is clear from Fig.~\ref{f:plots}, the allowed mass range of $\Delta_{R}^\pm$ from diphoton excess lies between $\sim$ 400 GeV to 540 GeV. We take this bound into account and take $m_{\Delta_{R}^\pm}=530$ GeV, $y_\chi=1.25$ and compute the relic density. As shown in Fig.~\ref{relic:plots} that the correct relic density is obtained for a dark matter mass 200 GeV.
As a result of dominant coupling of dark matter with $\Delta^\pm_R$, DM annihilates into leptons, evading the stringent bounds from its direct detection search \cite{Aprile2012,Akerib2014}.

\subsection{Muon ($g-2$)}
\label{sbsc:mugminus2}
There exists a $3.6\sigma$ discrepancy between SM prediction and experimental value of muon $(g-2)$ \cite{Bennett2006,Bennett2009}. 
\begin{figure}[t!]
 \begin{center}
 \includegraphics[scale=0.55]{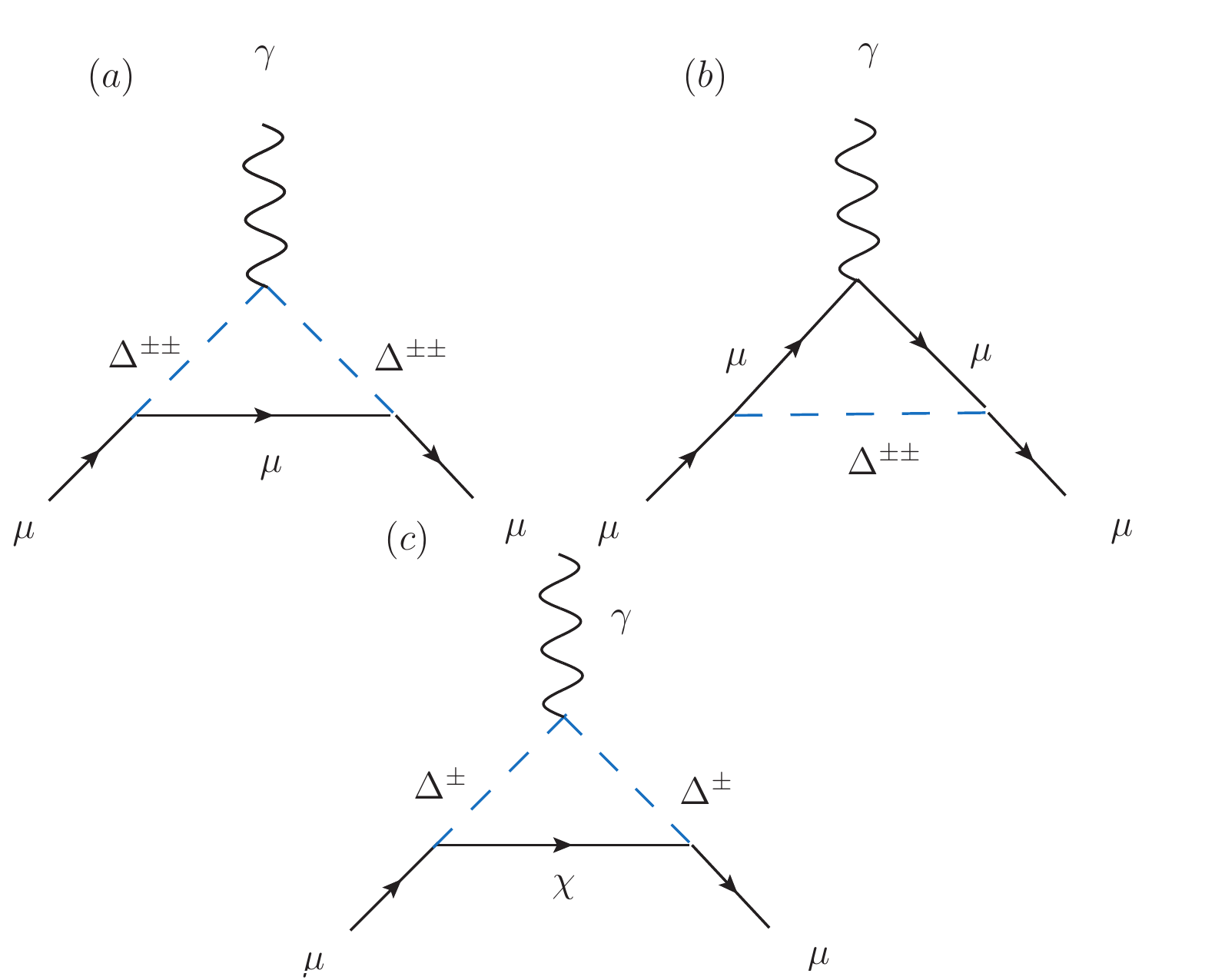}
\caption{\it Dominant Feynman diagrams of doubly (a,b) and  singly (c) charged triplet scalar loops contributing to muon $(g-2)$.}
\label{fig:mm}
 \end{center}
\end{figure}
 In DLRM \cite{Khalil:2009nb}, there exist additional gauge bosons and charged triplet scalars which 
 give contributions to the muon magnetic moment. As we mentioned before, there exist stringent bounds on the masses of $SU(2)_R$ gauge bosons ($W^\pm_R, Z_R$) from LHC, so the contributions of heavy gauge bosons to muon $(g-2)$ will be small in comparison to the charged scalars.  
 The relevant interactions term of muon $(g-2)$ are $\psi_R\psi_R\Delta_R$ and $\psi_L\psi_L\Delta_L$. But $\psi_L\psi_L\Delta_L$ gives the mass to the neutrino and so have a small Yukawa coupling, whereas $\psi_R\psi_R\Delta_R$ does not have any restriction like this. So, we take into account only the contribution from 
 $\Delta_R^\pm, \Delta_R^{\pm\pm}$ loops, as shown in Fig.~\ref{fig:mm}.\\
 The contribution from the doubly charged triplet higgs (as shown in Fig.~\ref{fig:mm}(a)-\ref{fig:mm}(b)) is given 
 by \cite{Leveille1978},
 \begin{align}\label{chc}\nonumber
 [\Delta a_\mu]_{\Delta^{\pm\pm}} = 4\times \bigg[\frac{2 m^2_\mu}{8\pi^2} \int^1_0 &dy \frac{f^2_{\mu s}(y^3-y)
 + f^2_{\mu p}(y^3-2y^2+y)}{m^2_\mu ~(y^2-2y+1) + m^2_{\Delta^{\pm\pm}}y}\\
 & - \frac{m^2_\mu}{8\pi^2} \int^1_0 dy \frac{f^2_{\mu s}(2y^2-y^3) - f^2_{\mu p} y^3}
 {m^2_\mu ~y^2 + m^2_{\Delta^{\pm\pm}}(1-y)}\bigg]
 \end{align}
where $f_{\mu s}$ and $f_{\mu p}$ are the scalar and pseudoscalar couplings of charged triplet higgs with the muon respectively. 
As a result of the interaction term ($\psi_R\psi_R\Delta_R$) which has two identical fields, the symmetry factor of 4 is appeared in Eq.~\ref{chc}.
The contribution from singly charged triplet higgs $(\Delta^\pm_R)$, which is shown in diagram \ref{fig:mm}(c), 
 given as \cite{Leveille1978},
 \begin{equation}\label{chc1}
 [\Delta a_\mu]_{\Delta^{\pm}}= \frac{m^2_\mu}{8\pi^2} \int^1_0 dy \frac{f^2_{\mu s}(y^3-y^2 +\frac{m_\chi}{m_\mu}
 (y^2-y))+f^2_{\mu p}(y^3-y^2 -\frac{m_\chi}{m_\mu}
 (y^2-y))}{m^2_\mu y^2 +(m^2_{\Delta^\pm}-m^2_\mu)y + m^2_\chi (1-y)}
 \end{equation}
 

The masses of the charged triplet scalars are fixed from the diphoton excess of LHC, which lies between $\sim 400-540$ GeV (see Fig.~\ref{f:plots}). We take $m_{\Delta^\pm}=m_{\Delta^{\pm\pm}} \sim 530$ GeV and use $f_{\mu s}=f_{\mu p}\simeq y_\chi = 1.25$ (this same value reproduces the correct relic density). We add all the contributions from Eq.~\ref{chc} and Eq.~\ref{chc1}, and finally obtain,  
\begin{equation}
 \Delta a_\mu = 2.99 \times 10^{-9},
\end{equation}
which is in agreement with the experimental result \cite{Bennett2006,Bennett2009} within $1\sigma$. The connection between the dark matter and muon $(g-2)$ in DLRM is also explored in \cite{Basak2015} with details.

\section{Summary and Conclusions}
\label{sec:concl} We explain the $750$ GeV diphoton resonance observed by ATLAS and CMS in the context of the existing dark left-right model. A global symmetry in DLRM ensures the stability of dark matter. The same global symmetry also ensures that the scalar particle identified as $750$ GeV resonance can dominantly couple to the standard model gluons and photons. This explain the large branching of $\gamma\gamma$ and $gg$ decay widths which is necessary to explain the diphoton cross-section. After fixing the parameters of the model to explain the diphoton cross-section, we compute the relic density of dark matter and muon $(g-2)$ in this model and we show that a common explanation for all three is possible in DLRM. This explanation predicts a dark matter mass of 200 GeV. 
It would be interesting to see if future observations of dark matter and LHC will corroborate or rule out this model.


\bibliographystyle{JHEP}
\bibliography{diphotonRef}

\providecommand{\href}[2]{#2}\begingroup\raggedright\begin{thebibliography}{10}

\bibitem{atl-dp}
{Tech. Rep. ATLAS-CONF-2015-081, CERN, Geneva (2015), \it
  http://cds.cern.ch/record/2114853}.

\bibitem{cms-dp}
{Tech. Rep. CMS-PAS-EXO-15-004, CERN, Geneva (2015), \it
  https://cds.cern.ch/record/2114808/files/EXO-15-004-pas.pdf}.

\bibitem{Agrawal2015}
P.~Agrawal, J.~Fan, B.~Heidenreich, M.~Reece, and M.~Strassler, {\it
  {Experimental Considerations Motivated by the Diphoton Excess at the LHC}},
  \href{http://arxiv.org/abs/1512.05775}{{\tt arXiv:1512.05775}}.

\bibitem{Aloni2015}
D.~Aloni, K.~Blum, A.~Dery, A.~Efrati, and Y.~Nir, {\it {On a possible large
  width 750 GeV diphoton resonance at ATLAS and CMS}},
  \href{http://arxiv.org/abs/1512.05778}{{\tt arXiv:1512.05778}}.

\bibitem{Angelescu2016}
A.~Angelescu, A.~Djouadi, and G.~Moreau, {\it {Scenarii for interpretations of
  the LHC diphoton excess: two Higgs doublets and vector-like quarks and
  leptons}},  {\em Phys. Lett.} {\bf B756} (2016) 126--132,
  [\href{http://arxiv.org/abs/1512.04921}{{\tt arXiv:1512.04921}}].

\bibitem{Backovic2015}
M.~Backovic, A.~Mariotti, and D.~Redigolo, {\it {Di-photon excess illuminates
  Dark Matter}},  \href{http://arxiv.org/abs/1512.04917}{{\tt
  arXiv:1512.04917}}.

\bibitem{Becirevic2015}
D.~Becirevic, E.~Bertuzzo, O.~Sumensari, and R.~Z. Funchal, {\it {Can the new
  resonance at LHC be a CP-Odd Higgs boson?}},
  \href{http://arxiv.org/abs/1512.05623}{{\tt arXiv:1512.05623}}.

\bibitem{Buttazzo2016}
D.~Buttazzo, A.~Greljo, and D.~Marzocca, {\it {Knocking on new physics’ door
  with a scalar resonance}},  {\em Eur. Phys. J.} {\bf C76} (2016), no.~3 116,
  [\href{http://arxiv.org/abs/1512.04929}{{\tt arXiv:1512.04929}}].

\bibitem{Cao2015}
Q.-H. Cao, Y.~Liu, K.-P. Xie, B.~Yan, and D.-M. Zhang, {\it {A Boost Test of
  Anomalous Diphoton Resonance at the LHC}},
  \href{http://arxiv.org/abs/1512.05542}{{\tt arXiv:1512.05542}}.

\bibitem{Chao2015}
W.~Chao, R.~Huo, and J.-H. Yu, {\it {The Minimal Scalar-Stealth Top
  Interpretation of the Diphoton Excess}},
  \href{http://arxiv.org/abs/1512.05738}{{\tt arXiv:1512.05738}}.

\bibitem{Csaki2016}
C.~Csáki, J.~Hubisz, and J.~Terning, {\it {Minimal model of a diphoton
  resonance: Production without gluon couplings}},  {\em Phys. Rev.} {\bf D93}
  (2016), no.~3 035002, [\href{http://arxiv.org/abs/1512.05776}{{\tt
  arXiv:1512.05776}}].

\bibitem{Curtin2016}
D.~Curtin and C.~B. Verhaaren, {\it {Quirky Explanations for the Diphoton
  Excess}},  {\em Phys. Rev.} {\bf D93} (2016), no.~5 055011,
  [\href{http://arxiv.org/abs/1512.05753}{{\tt arXiv:1512.05753}}].

\bibitem{Demidov2015}
S.~V. Demidov and D.~S. Gorbunov, {\it {On sgoldstino interpretation of the
  diphoton excess}},  \href{http://arxiv.org/abs/1512.05723}{{\tt
  arXiv:1512.05723}}.

\bibitem{DiChiara2015}
S.~Di~Chiara, L.~Marzola, and M.~Raidal, {\it {First interpretation of the 750
  GeV di-photon resonance at the LHC}},
  \href{http://arxiv.org/abs/1512.04939}{{\tt arXiv:1512.04939}}.

\bibitem{Dutta2015}
B.~Dutta, Y.~Gao, T.~Ghosh, I.~Gogoladze, and T.~Li, {\it {Interpretation of
  the diphoton excess at CMS and ATLAS}},
  \href{http://arxiv.org/abs/1512.05439}{{\tt arXiv:1512.05439}}.

\bibitem{Ellis2015}
J.~Ellis, S.~A.~R. Ellis, J.~Quevillon, V.~Sanz, and T.~You, {\it {On the
  Interpretation of a Possible $\sim 750$ GeV Particle Decaying into $\gamma
  \gamma$}},  \href{http://arxiv.org/abs/1512.05327}{{\tt arXiv:1512.05327}}.

\bibitem{Falkowski2015}
A.~Falkowski, O.~Slone, and T.~Volansky, {\it {Phenomenology of a 750 GeV
  Singlet}},  \href{http://arxiv.org/abs/1512.05777}{{\tt arXiv:1512.05777}}.

\bibitem{Franceschini2015}
R.~Franceschini, G.~F. Giudice, J.~F. Kamenik, M.~McCullough, A.~Pomarol,
  R.~Rattazzi, M.~Redi, F.~Riva, A.~Strumia, and R.~Torre, {\it {What is the
  gamma gamma resonance at 750 GeV?}},
  \href{http://arxiv.org/abs/1512.04933}{{\tt arXiv:1512.04933}}.

\bibitem{Gupta2015}
R.~S. Gupta, S.~Jäger, Y.~Kats, G.~Perez, and E.~Stamou, {\it {Interpreting a
  750 GeV Diphoton Resonance}},  \href{http://arxiv.org/abs/1512.05332}{{\tt
  arXiv:1512.05332}}.

\bibitem{Knapen2015}
S.~Knapen, T.~Melia, M.~Papucci, and K.~Zurek, {\it {Rays of light from the
  LHC}},  \href{http://arxiv.org/abs/1512.04928}{{\tt arXiv:1512.04928}}.

\bibitem{Kobakhidze2015}
A.~Kobakhidze, F.~Wang, L.~Wu, J.~M. Yang, and M.~Zhang, {\it {LHC diphoton
  excess explained as a heavy scalar in top-seesaw model}},
  \href{http://arxiv.org/abs/1512.05585}{{\tt arXiv:1512.05585}}.

\bibitem{Low2015}
M.~Low, A.~Tesi, and L.-T. Wang, {\it {A pseudoscalar decaying to photon pairs
  in the early LHC run 2 data}},  \href{http://arxiv.org/abs/1512.05328}{{\tt
  arXiv:1512.05328}}.

\bibitem{Mambrini2015}
Y.~Mambrini, G.~Arcadi, and A.~Djouadi, {\it {The LHC diphoton resonance and
  dark matter}},  \href{http://arxiv.org/abs/1512.04913}{{\tt
  arXiv:1512.04913}}.

\bibitem{Martinez2015}
R.~Martinez, F.~Ochoa, and C.~F. Sierra, {\it {Diphoton decay for a $750$ GeV
  scalar dark matter}},  \href{http://arxiv.org/abs/1512.05617}{{\tt
  arXiv:1512.05617}}.

\bibitem{McDermott2015}
S.~D. McDermott, P.~Meade, and H.~Ramani, {\it {Singlet Scalar Resonances and
  the Diphoton Excess}},  \href{http://arxiv.org/abs/1512.05326}{{\tt
  arXiv:1512.05326}}.

\bibitem{Petersson2015}
C.~Petersson and R.~Torre, {\it {The 750 GeV diphoton excess from the goldstino
  superpartner}},  \href{http://arxiv.org/abs/1512.05333}{{\tt
  arXiv:1512.05333}}.

\bibitem{Chakrabortty2015}
J.~Chakrabortty, A.~Choudhury, P.~Ghosh, S.~Mondal, and T.~Srivastava, {\it
  {Di-photon resonance around 750 GeV: shedding light on the theory
  underneath}},  \href{http://arxiv.org/abs/1512.05767}{{\tt
  arXiv:1512.05767}}.

\bibitem{Fichet2015}
S.~Fichet, G.~von Gersdorff, and C.~Royon, {\it {Scattering Light by Light at
  750 GeV at the LHC}},  \href{http://arxiv.org/abs/1512.05751}{{\tt
  arXiv:1512.05751}}.

\bibitem{Ahmed2015}
A.~Ahmed, B.~M. Dillon, B.~Grzadkowski, J.~F. Gunion, and Y.~Jiang, {\it
  {Higgs-radion interpretation of 750 GeV di-photon excess at the LHC}},
  \href{http://arxiv.org/abs/1512.05771}{{\tt arXiv:1512.05771}}.

\bibitem{Cox2015}
P.~Cox, A.~D. Medina, T.~S. Ray, and A.~Spray, {\it {Diphoton Excess at 750 GeV
  from a Radion in the Bulk-Higgs Scenario}},
  \href{http://arxiv.org/abs/1512.05618}{{\tt arXiv:1512.05618}}.

\bibitem{Chao2015a}
W.~Chao, {\it {Symmetries Behind the 750 GeV Diphoton Excess}},
  \href{http://arxiv.org/abs/1512.06297}{{\tt arXiv:1512.06297}}.

\bibitem{Bi2015}
X.-J. Bi, Q.-F. Xiang, P.-F. Yin, and Z.-H. Yu, {\it {The 750 GeV diphoton
  excess at the LHC and dark matter constraints}},
  \href{http://arxiv.org/abs/1512.06787}{{\tt arXiv:1512.06787}}.

\bibitem{Bardhan2015}
D.~Bardhan, D.~Bhatia, A.~Chakraborty, U.~Maitra, S.~Raychaudhuri, and
  T.~Samui, {\it {Radion Candidate for the LHC Diphoton Resonance}},
  \href{http://arxiv.org/abs/1512.06674}{{\tt arXiv:1512.06674}}.

\bibitem{Heckman2015}
J.~J. Heckman, {\it {750 GeV Diphotons from a D3-brane}},
  \href{http://arxiv.org/abs/1512.06773}{{\tt arXiv:1512.06773}}.

\bibitem{Barducci2015}
D.~Barducci, A.~Goudelis, S.~Kulkarni, and D.~Sengupta, {\it {One jet to rule
  them all: monojet constraints and invisible decays of a 750 GeV diphoton
  resonance}},  \href{http://arxiv.org/abs/1512.06842}{{\tt arXiv:1512.06842}}.

\bibitem{Cao2015a}
J.~Cao, C.~Han, L.~Shang, W.~Su, J.~M. Yang, and Y.~Zhang, {\it {Interpreting
  the 750 GeV diphoton excess by the singlet extension of the Manohar-Wise
  Model}},  \href{http://arxiv.org/abs/1512.06728}{{\tt arXiv:1512.06728}}.

\bibitem{Ding2015}
R.~Ding, L.~Huang, T.~Li, and B.~Zhu, {\it {Interpreting $750$ GeV Diphoton
  Excess with R-parity Violation Supersymmetry}},
  \href{http://arxiv.org/abs/1512.06560}{{\tt arXiv:1512.06560}}.

\bibitem{Han2015b}
H.~Han, S.~Wang, and S.~Zheng, {\it {Scalar Explanation of Diphoton Excess at
  LHC}},  \href{http://arxiv.org/abs/1512.06562}{{\tt arXiv:1512.06562}}.

\bibitem{Han2015a}
X.-F. Han and L.~Wang, {\it {Implication of the 750 GeV diphoton resonance on
  two-Higgs-doublet model and its extensions with Higgs field}},
  \href{http://arxiv.org/abs/1512.06587}{{\tt arXiv:1512.06587}}.

\bibitem{Alves2015}
A.~Alves, A.~G. Dias, and K.~Sinha, {\it {The 750 GeV $S$-cion: Where else
  should we look for it?}},  \href{http://arxiv.org/abs/1512.06091}{{\tt
  arXiv:1512.06091}}.

\bibitem{Antipin2015}
O.~Antipin, M.~Mojaza, and F.~Sannino, {\it {A natural Coleman-Weinberg theory
  explains the diphoton excess}},  \href{http://arxiv.org/abs/1512.06708}{{\tt
  arXiv:1512.06708}}.

\bibitem{Bai2015}
Y.~Bai, J.~Berger, and R.~Lu, {\it {A 750 GeV Dark Pion: Cousin of a Dark
  G-parity-odd WIMP}},  \href{http://arxiv.org/abs/1512.05779}{{\tt
  arXiv:1512.05779}}.

\bibitem{Bellazzini2015}
B.~Bellazzini, R.~Franceschini, F.~Sala, and J.~Serra, {\it {Goldstones in
  Diphotons}},  \href{http://arxiv.org/abs/1512.05330}{{\tt arXiv:1512.05330}}.

\bibitem{Bian2015}
L.~Bian, N.~Chen, D.~Liu, and J.~Shu, {\it {A hidden confining world on the 750
  GeV diphoton excess}},  \href{http://arxiv.org/abs/1512.05759}{{\tt
  arXiv:1512.05759}}.

\bibitem{Harigaya2015}
K.~Harigaya and Y.~Nomura, {\it {Composite Models for the 750 GeV Diphoton
  Excess}},  \href{http://arxiv.org/abs/1512.04850}{{\tt arXiv:1512.04850}}.

\bibitem{Matsuzaki2015}
S.~Matsuzaki and K.~Yamawaki, {\it {750 GeV Diphoton Signal from One-Family
  Walking Technipion}},  \href{http://arxiv.org/abs/1512.05564}{{\tt
  arXiv:1512.05564}}.

\bibitem{Nakai2015}
Y.~Nakai, R.~Sato, and K.~Tobioka, {\it {Footprints of New Strong Dynamics via
  Anomaly}},  \href{http://arxiv.org/abs/1512.04924}{{\tt arXiv:1512.04924}}.

\bibitem{No2015}
J.~M. No, V.~Sanz, and J.~Setford, {\it {See-Saw Composite Higgses at the LHC:
  Linking Naturalness to the $750$ GeV Di-Photon Resonance}},
  \href{http://arxiv.org/abs/1512.05700}{{\tt arXiv:1512.05700}}.

\bibitem{Higaki2016}
T.~Higaki, K.~S. Jeong, N.~Kitajima, and F.~Takahashi, {\it {The QCD Axion from
  Aligned Axions and Diphoton Excess}},  {\em Phys. Lett.} {\bf B755} (2016)
  13--16, [\href{http://arxiv.org/abs/1512.05295}{{\tt arXiv:1512.05295}}].

\bibitem{Molinaro2015}
E.~Molinaro, F.~Sannino, and N.~Vignaroli, {\it {Strong dynamics or axion
  origin of the diphoton excess}},  \href{http://arxiv.org/abs/1512.05334}{{\tt
  arXiv:1512.05334}}.

\bibitem{Pilaftsis2016}
A.~Pilaftsis, {\it {Diphoton Signatures from Heavy Axion Decays at the CERN
  Large Hadron Collider}},  {\em Phys. Rev.} {\bf D93} (2016), no.~1 015017,
  [\href{http://arxiv.org/abs/1512.04931}{{\tt arXiv:1512.04931}}].

\bibitem{Arun2015}
M.~T. Arun and P.~Saha, {\it {Gravitons in multiply warped scenarios - at 750
  GeV and beyond}},  \href{http://arxiv.org/abs/1512.06335}{{\tt
  arXiv:1512.06335}}.

\bibitem{Han2015}
H.~Han, S.~Wang, and S.~Zheng, {\it {Dark Matter Theories in the Light of
  Diphoton Excess}},  \href{http://arxiv.org/abs/1512.07992}{{\tt
  arXiv:1512.07992}}.

\bibitem{Cho2015}
W.~S. Cho, D.~Kim, K.~Kong, S.~H. Lim, K.~T. Matchev, J.-C. Park, and M.~Park,
  {\it {The 750 GeV Diphoton Excess May Not Imply a 750 GeV Resonance}},
  \href{http://arxiv.org/abs/1512.06824}{{\tt arXiv:1512.06824}}.

\bibitem{Huang2015}
F.~P. Huang, C.~S. Li, Z.~L. Liu, and Y.~Wang, {\it {750 GeV Diphoton Excess
  from Cascade Decay}},  \href{http://arxiv.org/abs/1512.06732}{{\tt
  arXiv:1512.06732}}.

\bibitem{Dhuria2015}
M.~Dhuria and G.~Goswami, {\it {Perturbativity, vacuum stability and inflation
  in the light of 750 GeV diphoton excess}},
  \href{http://arxiv.org/abs/1512.06782}{{\tt arXiv:1512.06782}}.

\bibitem{Hamada2015}
Y.~Hamada, T.~Noumi, S.~Sun, and G.~Shiu, {\it {An O(750) GeV Resonance and
  Inflation}},  \href{http://arxiv.org/abs/1512.08984}{{\tt arXiv:1512.08984}}.

\bibitem{Salvio2016}
A.~Salvio and A.~Mazumdar, {\it {Higgs Stability and the 750 GeV Diphoton
  Excess}},  {\em Phys. Lett.} {\bf B755} (2016) 469--474,
  [\href{http://arxiv.org/abs/1512.08184}{{\tt arXiv:1512.08184}}].

\bibitem{Jaeckel2013}
J.~Jaeckel, M.~Jankowiak, and M.~Spannowsky, {\it {LHC probes the hidden
  sector}},  {\em Phys. Dark Univ.} {\bf 2} (2013) 111--117,
  [\href{http://arxiv.org/abs/1212.3620}{{\tt arXiv:1212.3620}}].

\bibitem{Staub2008}
F.~Staub, {\it {SARAH}},  \href{http://arxiv.org/abs/0806.0538}{{\tt
  arXiv:0806.0538}}.

\bibitem{Staub2014}
F.~Staub, {\it {SARAH 4 : A tool for (not only SUSY) model builders}},  {\em
  Comput. Phys. Commun.} {\bf 185} (2014) 1773--1790,
  [\href{http://arxiv.org/abs/1309.7223}{{\tt arXiv:1309.7223}}].

\bibitem{Staub2016}
F.~Staub et~al., {\it {Precision tools and models to narrow in on the 750 GeV
  diphoton resonance}},  \href{http://arxiv.org/abs/1602.05581}{{\tt
  arXiv:1602.05581}}.

\bibitem{Khalil:2009nb}
S.~Khalil, H.-S. Lee, and E.~Ma, {\it {Generalized Lepton Number and Dark
  Left-Right Gauge Model}},  {\em Phys. Rev.} {\bf D79} (2009) 041701,
  [\href{http://arxiv.org/abs/0901.0981}{{\tt arXiv:0901.0981}}].

\bibitem{Aranda2010}
A.~Aranda, J.~Lorenzo Diaz-Cruz, J.~Hernandez-Sanchez, and E.~Ma, {\it {Dark
  Left-Right Gauge Model: SU(2)$_R$ Phenomenology}},  {\em Phys. Rev.} {\bf
  D81} (2010) 075010, [\href{http://arxiv.org/abs/1001.4057}{{\tt
  arXiv:1001.4057}}].

\bibitem{Basak2015}
T.~Basak, S.~Mohanty, and G.~Tomar, {\it {Explaining AMS-02 positron excess and
  muon anomalous magnetic moment in dark left-right gauge model}},
  \href{http://arxiv.org/abs/1501.06193}{{\tt arXiv:1501.06193}}.

\bibitem{Chatrchyan2013}
{\bf CMS} Collaboration, S.~Chatrchyan et~al., {\it {Search for heavy narrow
  dilepton resonances in $pp$ collisions at $\sqrt{s}=7$ TeV and $\sqrt{s}=8$
  TeV}},  {\em Phys. Lett.} {\bf B720} (2013) 63--82,
  [\href{http://arxiv.org/abs/1212.6175}{{\tt arXiv:1212.6175}}].

\bibitem{cmsl}
{CMS collaboration, \it
  https://cms-physics.web.cern.ch/cms-physics/public/HIG-12-005-pas.pdf.}

\bibitem{Aad2012}
{\bf ATLAS} Collaboration, G.~Aad et~al., {\it {Search for doubly-charged Higgs
  bosons in like-sign dilepton final states at $\sqrt{s}=7$ TeV with the ATLAS
  detector}},  {\em Eur. Phys. J.} {\bf C72} (2012) 2244,
  [\href{http://arxiv.org/abs/1210.5070}{{\tt arXiv:1210.5070}}].

\bibitem{Chatrchyan2012}
{\bf CMS} Collaboration, S.~Chatrchyan et~al., {\it {Search for a light charged
  Higgs boson in top quark decays in $pp$ collisions at $\sqrt{s}=7$ TeV}},
  {\em JHEP} {\bf 07} (2012) 143, [\href{http://arxiv.org/abs/1205.5736}{{\tt
  arXiv:1205.5736}}].

\bibitem{atlass}
{ATLAS collaboration, \it
  http://cds.cern.ch/record/1595533/files/ATLAS-CONF-2013-090.pdf.}

\bibitem{Ade2014}
{\bf Planck} Collaboration, P.~A.~R. Ade et~al., {\it {Planck 2013 results.
  XVI. Cosmological parameters}},  {\em Astron. Astrophys.} {\bf 571} (2014)
  A16, [\href{http://arxiv.org/abs/1303.5076}{{\tt arXiv:1303.5076}}].

\bibitem{Ellis1998}
J.~R. Ellis, T.~Falk, and K.~A. Olive, {\it {Neutralino - Stau coannihilation
  and the cosmological upper limit on the mass of the lightest supersymmetric
  particle}},  {\em Phys. Lett.} {\bf B444} (1998) 367--372,
  [\href{http://arxiv.org/abs/hep-ph/9810360}{{\tt hep-ph/9810360}}].

\bibitem{Nihei2002}
T.~Nihei, L.~Roszkowski, and R.~Ruiz~de Austri, {\it {Exact cross-sections for
  the neutralino slepton coannihilation}},  {\em JHEP} {\bf 07} (2002) 024,
  [\href{http://arxiv.org/abs/hep-ph/0206266}{{\tt hep-ph/0206266}}].

\bibitem{Cao2009}
Q.-H. Cao, E.~Ma, and G.~Shaughnessy, {\it {Dark Matter: The Leptonic
  Connection}},  {\em Phys. Lett.} {\bf B673} (2009) 152--155,
  [\href{http://arxiv.org/abs/0901.1334}{{\tt arXiv:0901.1334}}].

\bibitem{Aprile2012}
{\bf XENON100} Collaboration, E.~Aprile et~al., {\it {Dark Matter Results from
  225 Live Days of XENON100 Data}},  {\em Phys. Rev. Lett.} {\bf 109} (2012)
  181301, [\href{http://arxiv.org/abs/1207.5988}{{\tt arXiv:1207.5988}}].

\bibitem{Akerib2014}
{\bf LUX} Collaboration, D.~S. Akerib et~al., {\it {First results from the LUX
  dark matter experiment at the Sanford Underground Research Facility}},  {\em
  Phys. Rev. Lett.} {\bf 112} (2014) 091303,
  [\href{http://arxiv.org/abs/1310.8214}{{\tt arXiv:1310.8214}}].

\bibitem{Bennett2006}
{\bf Muon g-2} Collaboration, G.~W. Bennett et~al., {\it {Final Report of the
  Muon E821 Anomalous Magnetic Moment Measurement at BNL}},  {\em Phys. Rev.}
  {\bf D73} (2006) 072003, [\href{http://arxiv.org/abs/hep-ex/0602035}{{\tt
  hep-ex/0602035}}].

\bibitem{Bennett2009}
{\bf Muon (g-2)} Collaboration, G.~W. Bennett et~al., {\it {An Improved Limit
  on the Muon Electric Dipole Moment}},  {\em Phys. Rev.} {\bf D80} (2009)
  052008, [\href{http://arxiv.org/abs/0811.1207}{{\tt arXiv:0811.1207}}].

\bibitem{Leveille1978}
J.~P. Leveille, {\it {The Second Order Weak Correction to (G-2) of the Muon in
  Arbitrary Gauge Models}},  {\em Nucl. Phys.} {\bf B137} (1978) 63.

\end{thebibliography}\endgroup
\end{document}